# Aggregation Patterns of Salt Crystalizing in Drying Colloidal Solvents

**Moutushi Dutta Choudhury[a], Sayanee Jana[a], Sruti Dutta[a] and Sujata Tarafdar[a*]**

[a] *Condensed Matter Physics Research Centre, Physics Department, Jadavpur University, Kolkata 700032, India*
[*] *E-mail: sujata_tarafdar@hotmail.com*

We report a study of the structure of droplets of colloidal gels containing dissolved sodium chloride. The components segregate and form intricate patterns. The salt crystallizes in fractal and multi-fractal dendritic forms which are determined by the material which forms the colloidal gel. Here potato starch, gelatine and carboxymethyl cellulose have been used. The substrate also plays a role in some cases. Photographs and micrographs at different level of magnification are shown.

## Introduction

This study demonstrates how a very simple experiment of drying different colloidal solutions containing sodium chloride provides a fascinating array of rich and diverse patterns. Various aspects of evaporation of sessile drops on solid surfaces have long been a topic of interest [1-3]. The shape and time development of the contact line, the contour of the three dimensional liquid-vapour surface, the convection currents set up in the fluid[4], segregation of different components[5] and other interesting features have been observed and reported. The present paper deals with evaporation and drying of three component aqueous solutions, one of which is sodium chloride, leading to segregation and crystallization of the salt. The interesting point is that the morphology of the host colloid controls the aggregation process and the crystal deposits range through; simple cubic shapes, to faceted cuboids, highly branched multi-fractal dendritic patterns to extremely intricate designs resembling Julia sets depending upon the solvent and other conditions such as water content, temperature and humidity. The drying droplets have been observed at different scales of magnification and photographed by camera (Nikon CoolPixL120), polarizing microscope and by SEM imaging, each scale revealing new features. The solid substrate on which the sample dries also affects the pattern and the adhesion of the dried film to the substrate. We have used two different solids as the substrate – glass and polypropylene(PP).

The subject is not of academic interest only, there are practical applications as well, e.g. drying paints and coatings must be designed to avoid segregation of components, biomedical science utilizes the process for pathological investigation of biological fluids.

## Materials and Methods

In the present paper we report work on (1) potato starch gel (with varying concentration), (2) gelatine and (3) Carboxymethyl cellulose as the colloidal host fluid and sodium chloride as the dissolved salt. For comparison we also observe dried drops of sodium chloride in water and drying of starch films without salt.

**Two component solution: Salt in water**
When a drop of a solution of sodium chloride in distilled water is left to dry at room temperature (30ºC), one observes the well-known coffee stain effect[1,2]. Salt crystals form around the outer boundary of the drop (Fig.1). When a colloidal solution is used instead of pure water, the result is strikingly different. We discuss the different cases below.

**Two component solution: Starch in water**
Drying a drop of starch solution without any salt produces a transparent film, which detaches easily from the PP substrate, but sticks to the glass. During drying the outer fluid-solid contact line remains pinned but the boundary of the still fluid portion recedes inward. This is shown in the series of successive photographs (Fig. 2).

### Experiments with three-component solutions

**Potato Starch Gel**:
Colloidal gels of different proportion of the three components (x(salt) of NaCl, x(ps) of potato starch, MW $(162.14)_n$ and 50 ml of distilled water) have been prepared. The chemicals are procured from Loba Chemie, Mumbai, x(salt) mol of NaCl is dissolved in the water and slightly heated. Then x(ps) g of potato starch are added and the mixture is heated to 90ºC and stirred to get a homogeneous gel. Stirring is stopped and the gel is allowed to cool for 1 h.

We show results for x(salt) = 0.01 mol, x(ps) = 0.5 g and for x(salt) = 0.15 mol, x(ps) = 0.5 g. A drop of a 0.5 ml is deposited on a solid substrate (glass or PP) and left to dry. The dried film, which has a diameter of about 1.4 cm is observed under different magnifications.

**Images observed by the naked eye**

The appearance of the dried drops are very striking. For lower salt concentration a transparent band of starch separates along the outer periphery. On moving inward a number of narrower concentric rings are to be seen. In the central region the salt crystalizes forming fractal dendritic patterns. The films are photographed by Nikon CoolpixL120.

As the salt concentration increases, the band of starch around the edge grows successively narrower and finally disappears. Now the whole circular patch is covered by salt aggregates with a prominent cubic symmetry, but surrounded by fine dendritic structures.

Changing the substrate from glass to polypropylene (PP) does not change the patterns much, but on PP the film detaches cleanly from the substrate.

Fig. 3 shows the drops for two different compositions, on different substrates as photographed in close-up

**Microscope images**

Micrographs taken with Leica DM750 reveal very intricate and interesting details of each of the distinct rings seen with the naked eye. Fig. 4 Shows the detailed texture of each region of the complete film shown in (A), which is mapped schematically in the centre. Region (1) is the outermost layer comprising exclusively of starch. The granules are seen in the micrograph. As one focusses inward regions (2-4) contain more and more salt. The thickness of these rings are shown magnified in the map for clarity. In the innermost circle (5) salt dendrites are seen (they look dark) on a background of starch (which appears white).

**SEM images**

SEM images with EDAX show structural details at higher magnification and the composition of each of the layers. Two regions are shown in figures5. The outermost layer clearly consists mostly of starch. The micrographs and SEM images both show the typical starch grains and EDAX shows strong lines corresponding to C and O, Na and Cl lines are very weak here. On entering into the circular rings the composition reverses. Here Na and Cl are abundant and C, O much weaker. The SEM image shows intricate formations of salt. The inner regions have salt aggregates on a starch background as shown also by the micrographs in figure 4.

**Gelatine and Salt**

The next set of experiments is done with gelatine in water as the colloidal medium. 0.03 mole of NaCl is dissolved in 50 cc of water, then 0.5 g gelatine is mixed with the solution and stirred for 1 h at 70ºC until a homogeneous gel forms.

As in the previous set, drops are allowed to dry on two different substrates. The dried films look very different from the potato starch films shown in Fig. 3. There is no clear segregation of the salt either along the edge of the drop, or towards the centre. Instead, cubic crystals of salt form more or less uniformly throughout the film (Fig.6). Here the appearance of the dried films looks similar on both surfaces.

**Microscopic images with Leica DM750**

Under a microscope we can see interesting details shown in the two square frames marked in the Fig. 6. Focussing on the large crystals shows a regular step-like structure of the crystal. This is similar to what the crystals in Fig.1 look like under the microscope. The region in between the large crystals, now shows beautiful patterns of aggregation of the salt. The structure on analysis is found to be multi-fractal, further work in this direction is in progress[6] and will be reported elsewhere.

The result for a slightly different composition is also shown in Fig. 7. Here 0.03 mole NaCl is taken with 2g gelatine and 50 cc water and the gel is prepared in the same way as described before. For this thicker gel the salt aggregation pattern has a strong resemblance to a Julia set.

*A possible scenario for multi-fractal growth.*

Geometric multi-fractals are formed by strongly inhomogeneous growth processes[7]. We suggest an inhomogeneous process which may lead to a formation as shown in Fig.7, this is illustrated schematically in figure 8. After formation of the large crystals, the surrounding region is left depleted of dissolved salt. The drying process continues, leaving a very thick viscous gel, through which the remaining salt ions move very slowly towards microscopic nucleating crystallites, which are also square in shape. The aggregation is driven by the concentration gradient, which is strongest at the corners of the cubic crystal. The remaining ions therefore tend to attach preferentially at the corners of the seed crystal. This process may lead to formation of aggregates like the one in Fig. 7.

**Carboxymethyl cellulose(CMC) and salt**

Our last set of samples contains CMC as the gel former. Here 0.01 mole of NaCl and 0.1 g CMC are stirred in 50 cc water for 1 h at room temperature (32 ºC).

**Photographs and micrographs**

Fig. 9 shows photographs of the films cast on glass and PP substrates in (A) and (B) respectively. Corresponding micrographs are shown in figures (C) and (D).
Unlike the two earlier samples, where a strongly directed growth of crystalline aggregates was observed, here the photographs (A,B) show a pattern more similar to Diffusion Limited Aggregation(DLA)[7]. Possibly, the effect of the gel is strong enough to compete with the interaction between the salt ions leading to rock-salt structure formation. The micrographs show a more directed growth at short ranges, but the pattern is not like the multi-fractals observed earlier.
The effect of the substrate seems to play a role here. The structure of the film on glass does not have a ring like crystalline border which is seen in the film cast on the PP substrate.

## Discussion and Conclusions

This simple experiment reveals a rich variety of pattern formation processes at work in an evaporating drop of colloid with an added salt. The coffee stain effect[1,2] is now well studied. It has also been shown that Marangoni effect[7] can suppress this effect. Electro-wetting under different conditions has been used to control the coffee stain effect[8]. Strong phase segregation in albumen and salt solution has been observed by Tarasevich and Pravoslavnova[5]. However, a satisfactory explanation for such phenomena is still lacking.

Our observation is that the width of the band of starch drying out first along the boundary for the potato starch–salt gels, depends on the details of the composition and also ambient humidity. In our experiments it is unlikely that Marangoni effects which produce very weak temperature gradients will have a significant role. Moreover, as the colloid material changes from potato starch to gelatine to CMC, the overall structure of the film as well as the detailed structure of the salt crystal aggregates changes remarkably. We may infer that the visco-elastic properties of the colloids[9] and their microstructure affect the migration and aggregation pattern as the salt moves through the thickening gel. The subject appears to be a promising area for future study [10], particularly as it has already been proved to be a useful diagnostic technique[11].

Further work to understand the drying process, which is itself an interesting topic of research[12] is necessary as well as a theoretical analysis of the segregation phenomena under different conditions. The formation of fractal and multi-fractal aggregates may be simulated through random walks or Laplacian growth processes similar to viscous fingering[13], driven in this case by the concentration gradient.

## Acknowledgements

MDC is grateful to the Alumni Association, JU for providing a research scholarship endowed by Shantanu Das. We thank Prof. S. P. Moulik for helpful suggestions and for providing the CMC sample. SEM imaging was done at Central Glass and Ceramic Research Institiute, Kolkata.

## Notes and references

Figure 1: A, B, C and D show the residues left on evaporating drops of salt solutions made from 0.01, 0.02, 0.03 and 0.10 mole NaCl in 50 cc de-ionised water. The salt crystals form along the edge only, demonstrating the well-known 'coffee stain' effect.

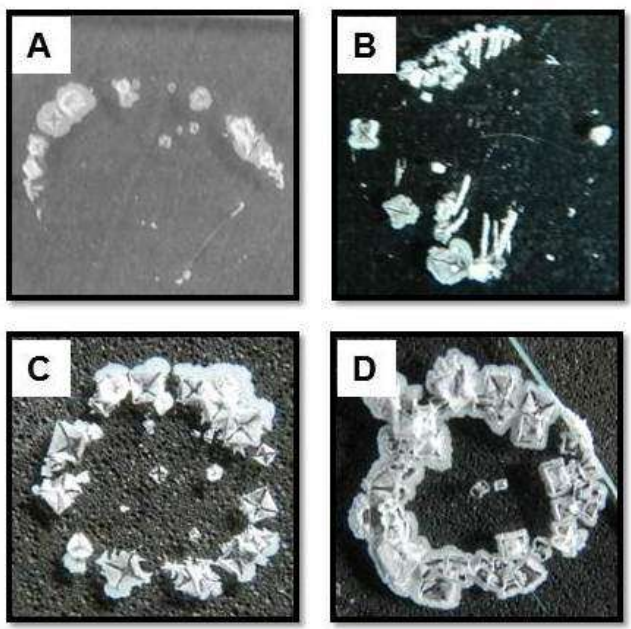

Figure 2: A, B, C and D show four stages during drying of a drop of potato starch gel with no salt. The initial three-phase contact line stays pinned to the substrate, but the central portion remains in the liquid state, while the dried starch forms a slowly widening band around the edge.

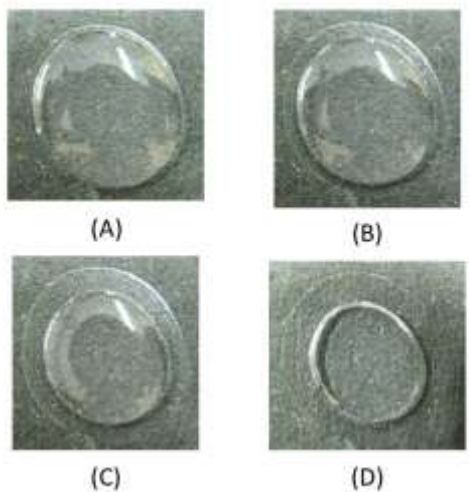

Figure 3: A and B show photographs of the two films with 0.01 mol NaCL, 0.5 g potato starch in 50 ml water on glass and PP. C and D show films with 0.15 mol NaCl, other components remaining same on glass and PP.

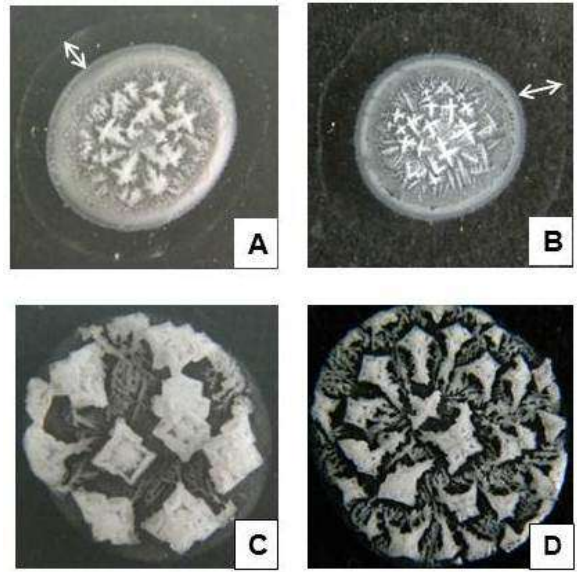

Micrographs of the different regions of the film containing potato starch, NaCl (0.15 g) and water. The schematic map below indicates the regions seen in the photograph of the film marked A.

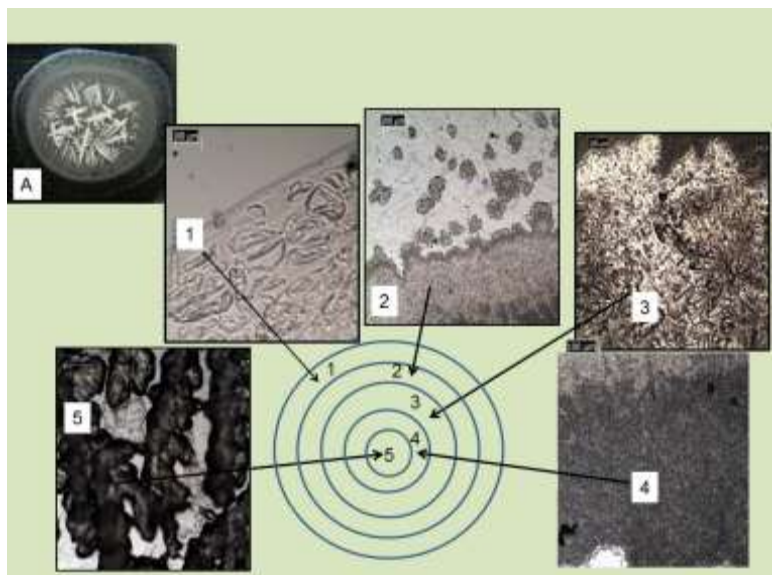

Figure 5. SEM and EDAX of the regions 1 and 3 in figure 4, show that the outermost band contains almost exclusively starch, while the salt aggregates near the central region.

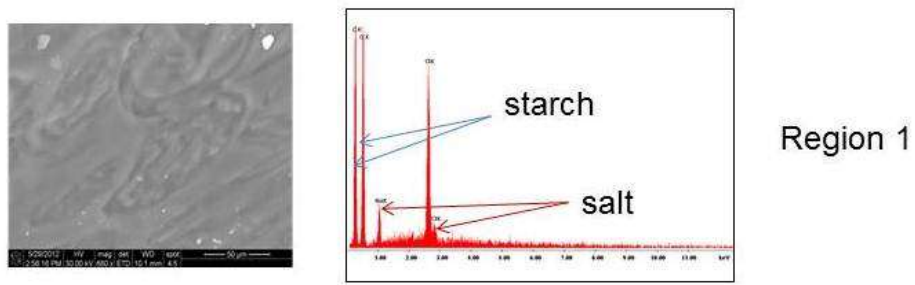

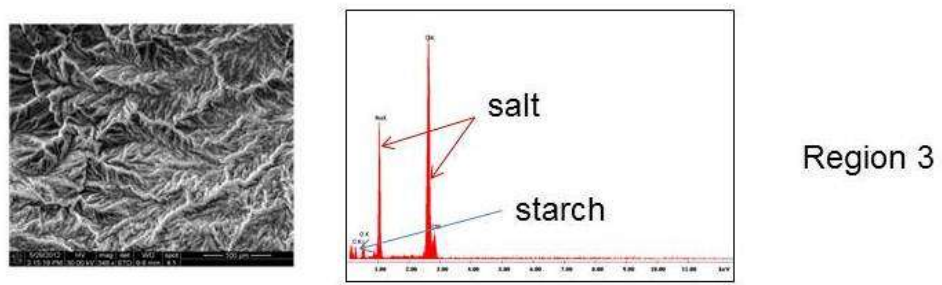

Figure 6: Film of NaCl in gelatine and water is shown in A. Here salt forms large crystals distributed more or less uniformly through the whole film. A micrograph of the large crystal is shown on the left below and a micrograph of the region in between large crystals is shown on the right. An intricate pattern of dendritic aggregates is seen.

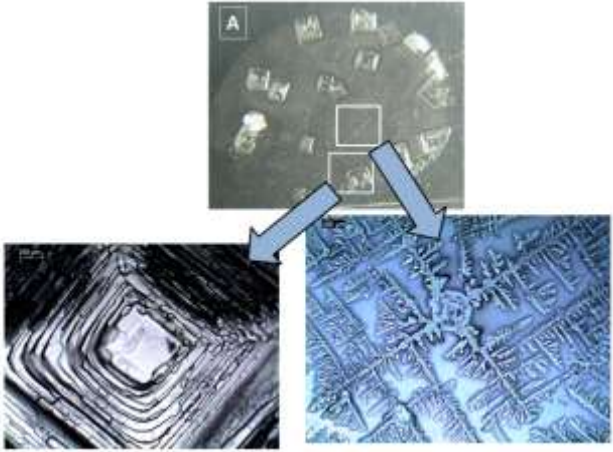

Figure 7. For higher concentration of NaCl in gelatine compared to figure 6, shows an aggregate very similar to a Julia set.

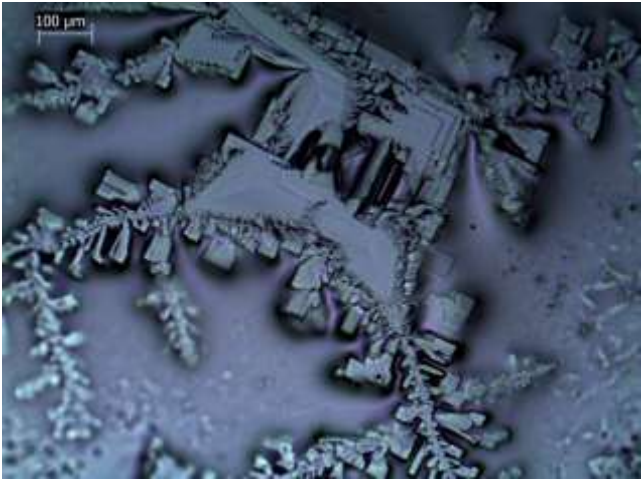

Figure 8. Schematic representation of a possible process leading to multi-fractal Julia set like aggregation as shown in figure 7. The sparsely distributed ions in the region between the large crystals are driven by a concentration gradient and attach to a cubic seed crystallite. The concentration gradient being large at the corners, growth proceeds preferentially from the corners of the seed.

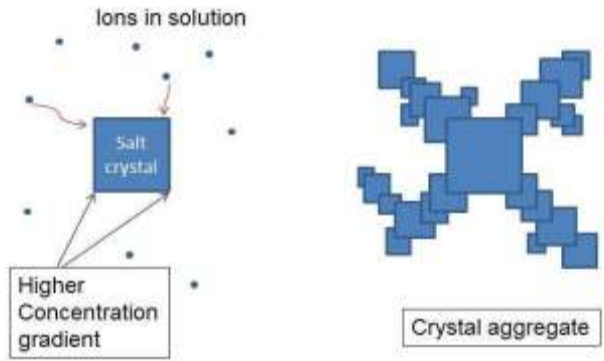

Fiigure 9. NaCl in a colloidal solution of CMC in water, shows different aggregation patterns on glass (A and C) and PP (B and D). A and B show the complete films photographed by Nikon CoolPix and C and D are magnified views of the aggregates through microscope.

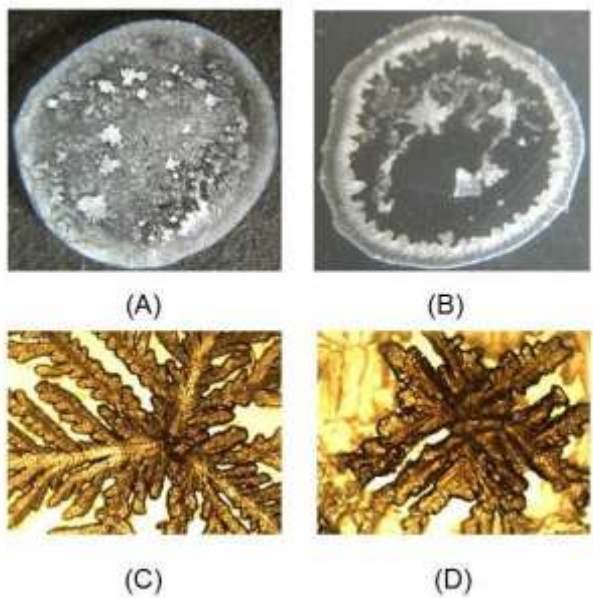